\documentclass[sigconf]{acmart}
\AtBeginDocument{%
  }

\copyrightyear{2026}
\acmYear{2026}
\setcopyright{cc}
\setcctype{by}
\acmConference[CUI '26]{ACM Conversational User Interfaces 2026}{July 21--24, 2026}{Bremen, Germany}
\acmBooktitle{ACM Conversational User Interfaces 2026 (CUI '26), July 21--24, 2026, Bremen, Germany}
\acmDOI{10.1145/3816046.3816314}
\acmISBN{979-8-4007-2741-2/2026/07}

\usepackage{xcolor}
\begin{document}

\title{Explanations as Dialogues: Toward Human-Centered Conversational Explainable AI}

\author{Niharika Mathur}
\email{nmathur35@gatech.edu}
\orcid{0000-0002-3969-7787}
\affiliation{%
  \institution{Georgia Institute of Technology}
  \city{Atlanta}
  \state{Georgia}
  \country{USA}
}

\author{Smit Desai}
\email{sm.desai@northeastern.edu}
\orcid{0000-0001-6983-8838}
\affiliation{%
  \institution{Northeastern University}
  \city{Boston}
  \state{Massachusetts}
  \country{USA}
}

\renewcommand{\shortauthors}{Mathur and Desai}
\renewcommand{\shorttitle}{Toward Human-Centered Conversational XAI}

\begin{abstract}
As AI systems become increasingly conversational, a gap emerges wherein explanations are studied as static artifacts, yet in practice, are experienced as dialogue. In this provocation, we argue that the conversational layer around an explanation is not incidental to its effectiveness, but a critical constituent. Drawing on three illustrative scenarios, we invite the CUI community to study explanations as interactive, conversational exchanges shaped by timing, tone, persona and conversational history, and introduce our vision for Human-Centered Conversational XAI (HC2XAI).
\end{abstract}



\maketitle

\section{Introduction}

\textcolor{black}{Over the past decade, Explainable AI (XAI) has emerged as a significant area of research, driven by growing concerns around algorithmic accountability \cite{chia2023emergence}, regulatory pressure \cite{gunning2019darpa}, and user trust \cite{thalpage2023unlocking, rosenbacke2024explainable}.}
The field has developed a diverse toolkit, from saliency maps, counterfactual explanations, feature importance scores, attention visualizations, to natural language summaries. In theory, AI explanations are system outputs that reveal the inner mechanisms of AI systems in order to provide reasoning or justification for their responses. Significant technical algorithms such as LIME \cite{knab2025lime} and SHAP \cite{kedar2024exploring} have given AI practitioners accessible methods for opening the “black-box” of complex AI models. More recently, large language models (LLMs) have enabled generating fluent natural-language explanations of AI behavior. Today, an AI system can articulate the reasoning behind a decision in grammatically sound prose in various languages and conversational personalities \cite{nighojkar2025giving, gavrilova2026conversing}. 

Despite these strides, the implicit assumption underlying much of explanatory progress, however, has remained stable: that a “good” explanation is an inherent property of content and faithfulness \cite{miller2019explanation, miller2023explainable}. As illustrated in Figure 1, a technically faithful explanation can be simultaneously conversationally inert and unable to adapt or invite follow-up from a user and their needs. The drill has been to obtain the correct information from the model, translate it into a legible modality and assume that user understanding will follow. Under this view, explanations act like artifacts that are produced, delivered and received. However, in recent years, something fundamental has shifted about the context in which that delivery happens. The AI systems that people interact with today are not dashboards or reports appended to a model output. They are \textit{conversational}. People can (1) ask why a system recommended one option over another, (2) interrupt a response mid-stream to request clarification, and (3) return days later, expecting it to remember everything said before. In each of these cases, the explanation does not by itself exist in isolation. It exists in the interaction through a voice, a conversational tone or a turn in an ongoing dialogue. It lands differently depending on who is asking for it, how they asked, what was asked before, their emotional state at the time of interaction, and what is at stake for them in this moment \cite{mathur2025sometimes, ehsan2024xai}.

HCI has made important strides in reorienting XAI towards centering user needs through emergence of Human-Centered XAI (HCXAI) \cite{ehsan2021expanding, liao2020questioning}. HCXAI has provided a counterargument for the assumption that technical fidelity alone is sufficient for AI explanations, instead insisting that explanations must be grounded in user goals, mental models and real-world decision contexts \cite{kim2023help}. While this disruption has been undeniably productive, even within HCXAI, the dominant mode of explanation under study remains largely visual and static: charts, highlight overlays, summary panels, structured templates, etc. \cite{kim2023help, ferguson2024explanation, gupta2024comparative}. Explanation is still an artifact that is shown to users rather than something that is said, and not something that is co-constructed in real time between a user and an AI system through conversational interaction.
\textcolor{black}{Recent work has begun to explore conversational approaches to XAI, including dialogue-based explanation systems and conversational interfaces for navigating AI outputs \cite{zhang2025conversational, jentzsch2019conversational, shen2023convxai}. However, much of this work still treats conversation primarily as a delivery mechanism for explanations. In contrast, we argue that the conversational layer itself, including timing, tone, repair, persona, pacing and interactional history, fundamentally shapes how explanations are interpreted and negotiated in practice, and necessitates a human-centered inquiry \cite{he2025conversational}.}

\begin{figure*}
    \centering
    \includegraphics[width=0.6\linewidth]{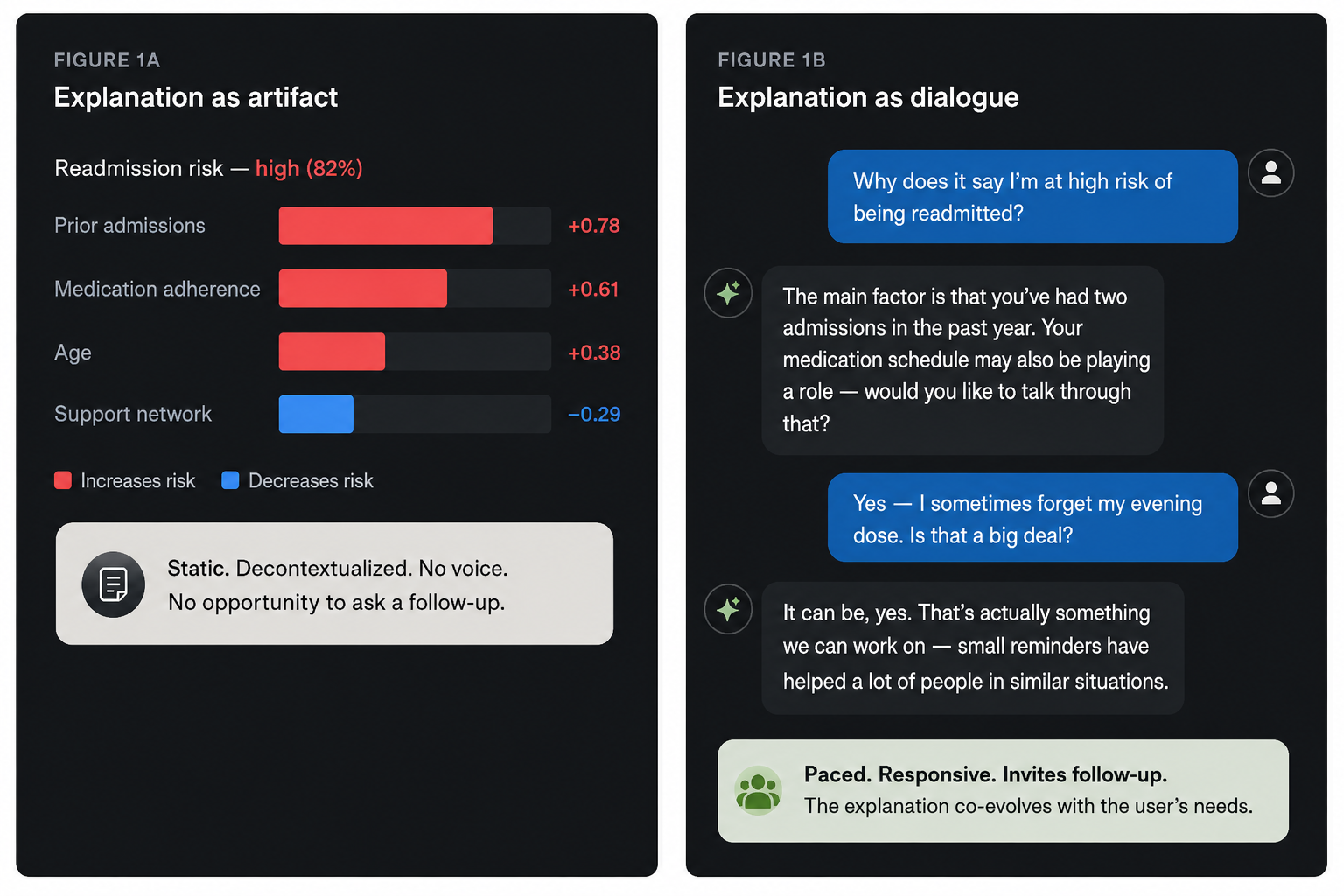}
    \caption{\textcolor{black}{A hospital readmission risk score rendered as a static SHAP-style feature importance chart (Fig. 1A) versus as a conversational exchange (Fig. 1B).}}
    \label{fig:one}
    \Description{This figure presents a case when a technically correct explanation may feel conversationally inert using the example of an AI-assisted hospital readmission risk score assessment. In Fig. 1A, the AI presents percent confidence scores for each feature that contributes to the final readmission risk assesment, while Fig. 1B shows the same information but in a conversational form with options for the user to ask follows up.}
\end{figure*}

In this provocation, here lies the gap that we want to highlight. As conversational AI becomes a primary interface through which many users encounter AI systems and, therefore, the primary medium through which these systems explain themselves, the field needs a framework or a starting point that takes the \textit{“conversational”} seriously, not just a presentation layer but a site of conversational sensemaking. Addressing this gap will likely require closer alignment between communities that have largely progressed in parallel, particularly those working on conversational user interfaces (CUIs) and those advancing explainability, to collectively explore what explanation-as-dialog might entail in practice. \textcolor{black}{We refer to this emerging space as} Human-Centered Conversational XAI, or HC2XAI, and offer this provocation as an opening vision towards articulating what it might involve.

\begin{figure*}
    \centering
    \includegraphics[width=0.8\linewidth]{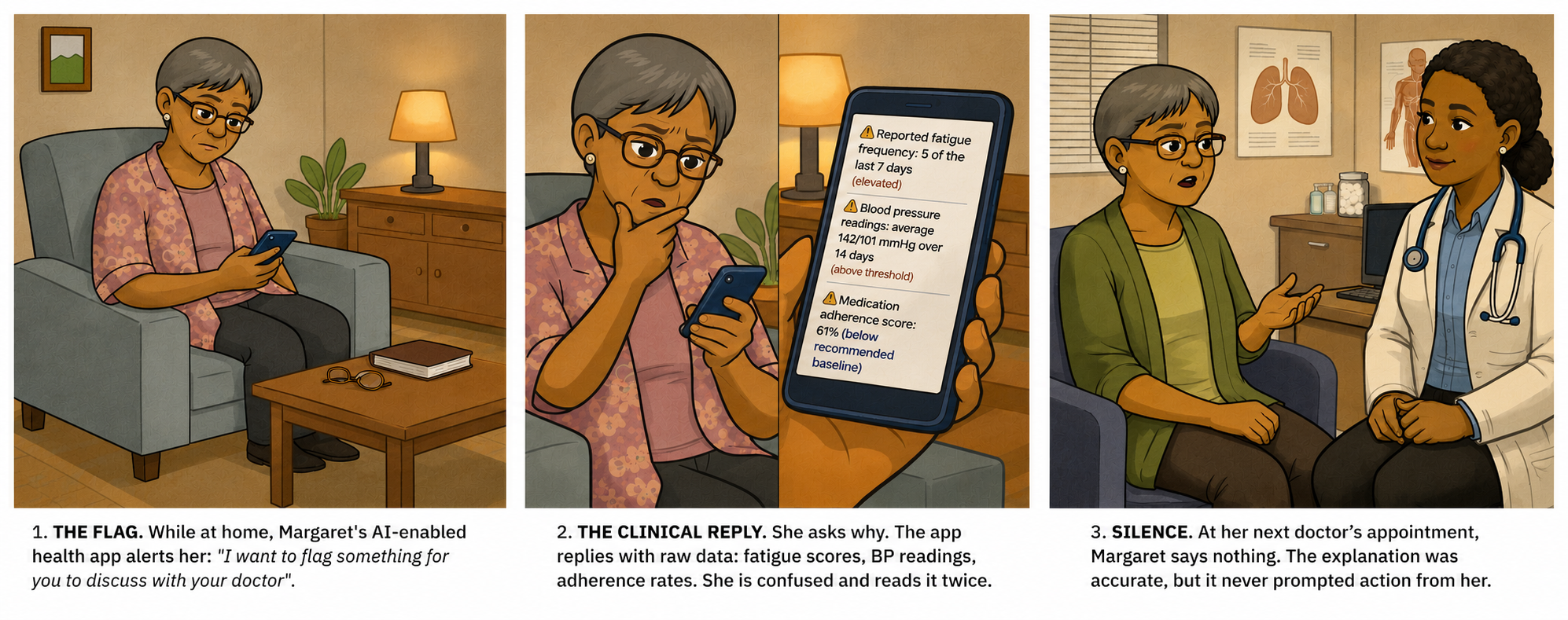}
    \caption{\textcolor{black}{Scenario A: An Older Adult Interacting with an AI-enabled Health Management App.}}
    \label{fig:two}
    \Description{A three-panel storyboard depicts an older woman named Margaret interacting with an AI-enabled health application. In the first panel, Margaret receives a notification asking her to discuss a health concern with her physician. In the second panel, the application presents technical data such as fatigue frequency, blood pressure readings, and medication adherence scores, leaving her visibly confused and uncertain. In the final panel, Margaret attends a medical appointment but remains silent, illustrating how clinically framed explanations may fail to encourage understanding or action despite being technically accurate.}
\end{figure*}

\section{Background: Where HCXAI Has Brought Us}

Before laying out the vision for HC2XAI, it is important to examine what HCXAI has accomplished. The shift from technical XAI to human-centered XAI marked a significant disciplinary move, reframing the central question from “is this explanation faithful to the model?” to “is this explanation useful to a person?” Research has shown that explanations performing well on fidelity benchmarks often fail to support effective human decision-making \cite{kulesza2013too, fok2024search}, giving rise to questions of “explainability for whom” \cite{gerlings2021explainable}. In parallel, attention to lay users, non-expert populations, and marginalized communities has foregrounded concerns around accessibility and equity \cite{mathur2023did}, including how explanations designed for expert audiences may exclude users with lower AI literacy \cite{long2023ai}, and how dominant formats assume familiarity with computational systems \cite{gregor1999explanations}. Early work on automated rationale generation further explored translating model outputs into natural language summaries more accessible to non-expert users \cite{ehsan2018rationalization, ehsan2019automated}, representing an initial step toward explanations as dialogues. For example, rather than presenting a bar chart indicating “\textit{prior admissions = +0.78},” such systems might instead generate: “\textit{Your past hospital visits were the strongest factor in this prediction}.”

And yet, much of the evaluation infrastructure in XAI continues to treat explanations as stimuli: users interact with an explanation and their responses are measured. While this marks a step forward, the paradigm remains largely one-shot and controlled \cite{schmid2022missing}. Even in longitudinal settings, evaluations tend to focus on how perceptions of explanation quality change over time, rather than how explanations unfold \textit{within} an ongoing interaction between a user and an AI \cite{gjoreski2024validate, zafari2024trust, long2025facilitating}. As a result, conversational dynamics such as turn-taking, repair, follow-up, tone, and personality remain largely invisible. This reflects the state of AI when much of HCXAI emerged, where interaction was primarily mediated through static, one-directional interfaces such as dashboards, decision panels, or summary reports. However, with the advent of generative AI and LLMs, a growing share of explanatory interactions now occurs within conversational systems \cite{yi2024revolutionizing, araujo2024speaking}. In such settings, the effectiveness of an explanation is shaped not only by \textit{what} is said, but \textit{how} it is said, \textit{when} it is said, and how it responds to the user in the moment. This raises a broader question: have our evaluation frameworks evolved alongside the systems they aim to study?. 

Within related research communities, conversation analysis offers a rich vocabulary for understanding how explanations function in talk as accounts, justifications, and repairs, and how their sequential placement shapes interaction \cite{blum2010explanations}. Dialogue systems research has long grappled with grounding and the establishment of common ground \cite{alikhani2020achieving, clark1989contributing}, while CUI work on persona, voice, and conversational style provides direct insight into how delivery shapes perception \cite{mathur2026wants, desai2025personas, namvarpour2025art, Rahman_Desai_2025, Desai_Chin_Wang_Cowan_Twidale_2025}. Emerging work on proactivity further raises questions about when and how systems should surface information, particularly across low- and high-stakes scenarios, and how such timing shapes user interpretation and action \cite{Zargham_Reicherts_Bonfert_Voelkel_Schoening_Malaka_Rogers_2022, Abulimiti_Pena_Alizadeh_Ahire_Candello_Desai_Edwards_He_Higgins_Jovane_et_al._2025}. In conjunction, this body of work points toward a broader view of explanation as not only a matter of content, but of timing, delivery, and interactional context. To establish a robust agenda for HC2XAI, we must bring these threads together and focus them on explanations as a conversational act.

\section{Scenarios: Explanations in the Wild}

To demonstrate what we mean by explanations as conversational acts, we present the following scenarios. These scenarios are not entirely speculative. They are composites of interactions already occurring in everyday lives, such as in consumer AI products, in assistive and health technology contexts, and in AI-integrated creative workflows. Our goal with these scenarios is to present them as design and research provocations in a narrative form.

\subsection{Scenario A: Margaret and the Health App.}

This scenario, adapted from studies examining how older adults interact with AI explanations \cite{mathur2025sometimes, li2023privacy}, follows Margaret, a 72-year-old living independently, who uses an AI-powered medication management app that her daughter helped set up after a recent hospital stay (Fig. 2). The app tracks her symptoms, flags anomalies, and provides plain-language summaries of her health patterns. One evening, it identifies an unusual pattern and states: \textit{"I want to flag something for you to discuss with your doctor."} When Margaret asks, \textit{"What did I say that worried you?"}, the system responds with a set of clinical metrics rendered in technical language: \textit{"Reported fatigue frequency: 5 of the last 7 days (elevated). Blood pressure readings: average 142/91 mmHg over 14 days (above threshold). Medication adherence score: 61\% (below recommended baseline)."} This response is conversationally distant from the kind of interaction she is accustomed to having with the assistant. She reads it twice, confused, puts her phone down, and does not bring it up at her next appointment. In Margaret’s case, the explanation is technically accurate, but conversationally inert; it does not meet her where she is, nor does it account for the possibility that she may be alone, in a hurry, or predisposed, as is common among older adults \cite{stoller1993interpretations}, to underreport concerns when information feels overwhelming or when she fears being dismissed. \textcolor{black}{Here, while the explanation may be too technical and that may be cause for confusion as well, but more importantly, our focus is on how the explanation is interactionally insensitive to Margaret's needs. The system delivers clinically dense information in a single turn without adapting to Margaret’s confusion, emotional state, prior conversational patterns or likelihood of needing clarification. Additionally, it also has access to weeks of interaction history with Margaret and a working model of how she communicates, and yet this context does not show up in how the explanation unfolds conversationally.}

In such a scenario, what would it mean to design that explanation conversationally? It could mean pacing it across multiple turns, inviting follow-up before proceeding, or calibrating the register to Margaret's demonstrated comprehension patterns: choices that emerging work suggests are meaningfully shaped by the personality dimensions of the delivering agent, such as agreeableness (e.g., \cite{mathur2026wants}), or conversational styles such as directness (e.g., \cite{Cox_Wester_Berkel_2026}).

\begin{figure*}
    \centering
    \includegraphics[width=0.8\linewidth]{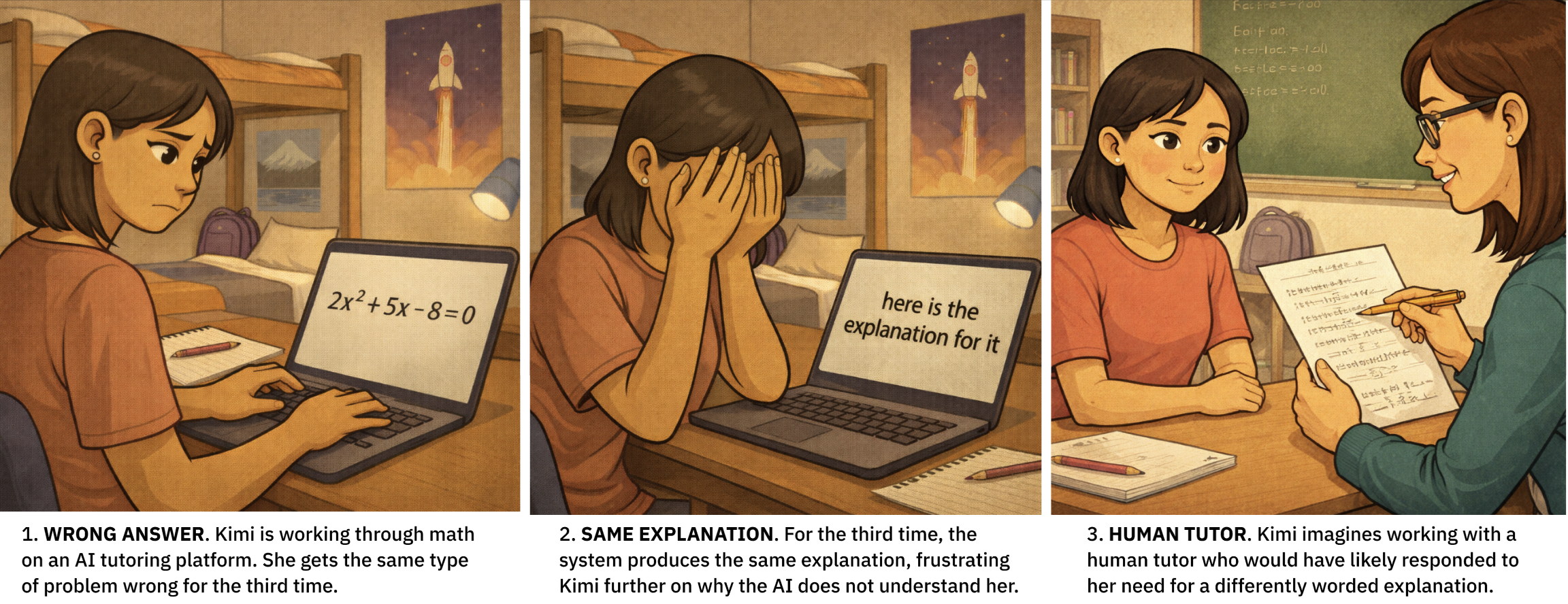}
    \caption{Scenario B: A college student interacting with an AI tutor.}
    \label{fig:three}
    \Description{A three-panel storyboard shows a young student named Kimi using an AI tutoring platform while solving a math problem. After repeatedly getting the same type of problem wrong, the system provides the same generic explanation again, causing visible frustration and emotional disengagement. Kimi covers her face in frustration as the repeated explanation appears on the screen. In the final panel, she imagines working with a human tutor who could adapt the explanation to her learning needs.}
\end{figure*}

\subsection{Scenario B: Kimi and the AI tutor.}
This scenario, adapted from recent research on the integration of AI tutoring systems in large courses \cite{Russell_Smith_George_Pratt_Fodale_Monk_Brummett_2025}, imagines Kimi, a second-year undergraduate student preparing for an exam using such a platform. She gets a multi-step algebra problem wrong, the same type of problem that she has got wrong twice before in the same session (Fig. 3). The system detects this and generates an explanation. For the third time, it produces the same response: \textit{"To solve for x, isolate variable by subtracting 3 from both sides. Then divide both sides by 2: x = 4"}. The explanation is accurate, and is also structurally identical to the two explanations it gave her before for the same problem. What it cannot see, however, is that Kimi is not making an arithmetic error. She is misreading which term to isolate first, a conceptual confusion that the same explanation might now be able to resolve. A human tutor in the same situation would not repeat themselves verbatim while explaining to Kimi in successive attempts. They would notice the pattern of errors, change the approach, ask what Kimi thinks went wrong, try a different strategy or analogy, or step back to check her understanding of a prerequisite concept. This demonstrates that the conversational competence of a skilled tutor lies in more than \textit{what} they explain, but also in how they \textit{“read the room”} \cite{mathur2025sometimes} and how they adjust in real time to what a student's responses reveal about their understanding. However, this context informed only what was explained, not how it was delivered. Here, recent CUI work on Mutual Theory of Mind points toward how such interactions could be more responsively and dynamically shaped \cite{Wang_Saha_Gregori_Joyner_Goel_2021}. Building on this, HC2XAI might ask: how can explanation systems draw on conversational history not just to personalize content, but to adapt the \textit{tone} and \textit{timing} of the explanation itself?

\subsection{Scenario C: David and a travel recommendation.}

This scenario, derived from \cite{Rahman_Desai_2025}, follows David, who is planning a weekend trip and has been chatting with an LLM travel assistant, noting that he prefers the outdoors, is on a budget, and is traveling with a child (Fig. 4). When the assistant recommends a destination and David asks why, it draws on prior turns: \textit{"Earlier you mentioned you were traveling with a child and preferred outdoor activities, so I prioritized family-friendly parks over urban destinations."} This is not an explicitly designed feature, but an emergent property of the interaction. Functionally, it reflects conversational XAI in action, where dialogue history is used to situate the explanation in a way that feels personally relevant. In this sense, it exemplifies a version of HC2XAI as we envision it.

At the same time, the explanation is produced without clear design intent or evaluation of whether it actually supports understanding. While plausible and useful, it may also create an unwarranted sense of coherence. The system references prior preferences, but does not indicate whether they meaningfully shaped the recommendation or were assembled into a \textit{post-hoc} rationale. As a result, David cannot distinguish between a faithful explanation and a conversationally constructed justification. This ambiguity directly affects trust calibration. The fluency and contextual alignment of the explanation may encourage over-trust without supporting verification, while misaligned explanations in other contexts may lead to under-trust \cite{Dubiel_Daronnat_Leiva_2022, mathur2025sometimes} and eventual disuse \cite{Luger_Sellen_2016}. In both cases, the issue is not simply correctness, but whether the explanation supports appropriately calibrated trust, a key driver of sustained engagement in CUIs \cite{Cowan_Pantidi_Coyle_Morrissey_Clarke_Al-Shehri_Earley_Bandeira_2017}.

Rather than an edge case, this highlights a broader challenge for conversational explanations: how to leverage interactional context without over-signaling coherence or obscuring uncertainty. We argue that the CUI community is well-positioned to study and design for these dynamics, rather than allowing them to emerge as byproducts of personalization.

\begin{comment}
Consider what happens when a user asks an LLM-powered chatbot to explain why it made a particular recommendation. Increasingly, the systems do not treat this as a fresh prompt and their response references something the user said several turns earlier in the conversation: \textit{"Earlier you mentioned you were trying to minimize costs, so I weighted..."} This is not a designed feature, but an emergent property of the current architecture. It is, functionally, conversational XAI in action: the system is using the dialogue history to situate and contextualize the explanation in a way that is relevant to the user. 

What is notable about this is precisely that it is happening without intentional design, without theoretical grounding and perhaps even without any evaluation of whether it is actually helping users. While it is sometimes surely useful, it may create false impressions of coherence or reasoning depth at other times, a kind of a conversational ELIZA effect \cite{}, where the appearance of contextual responsiveness is mistaken for genuine understanding \cite{}, without the user feeling the need to verify its accuracy (i.e., did I actually say that before?). In our view, the CUI research community is well-positioned to study this empirically and to design for it deliberately, rather than letting it emerge as a side effect of personalized AI interactions.
\end{comment}

\begin{figure*}
    \centering
    \includegraphics[width=0.8\linewidth]{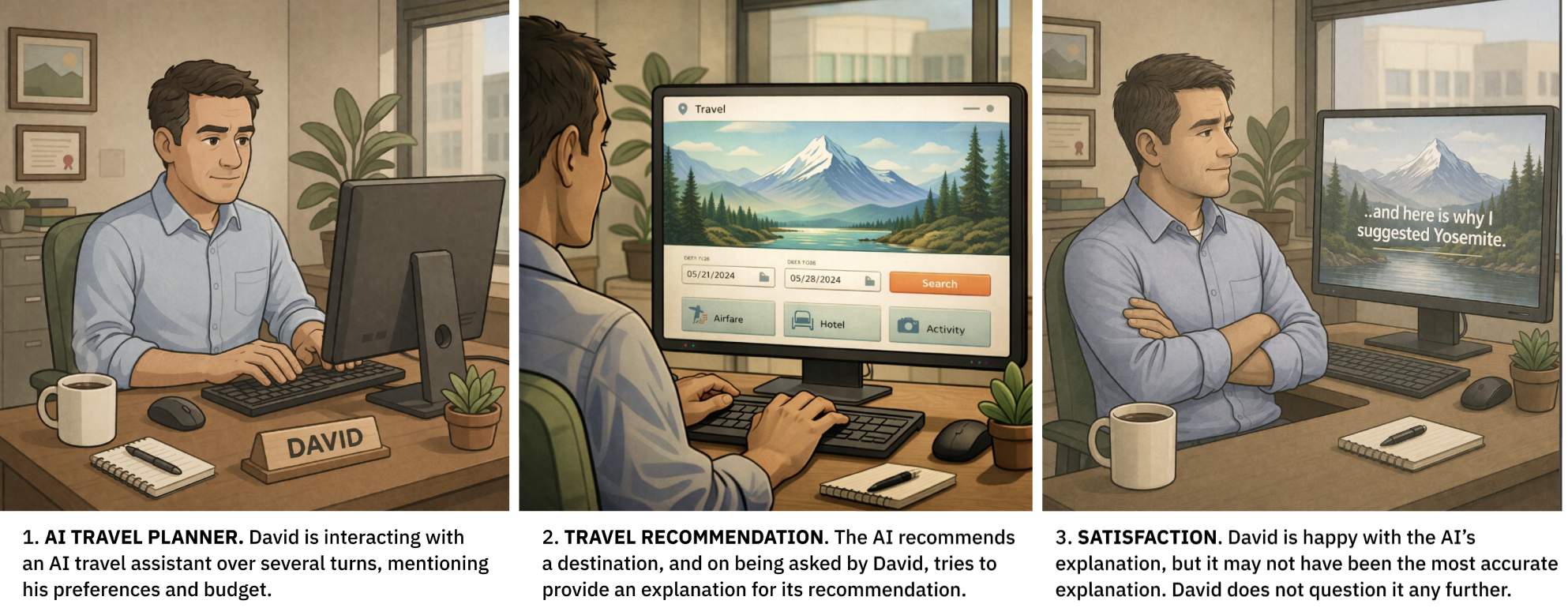}
    \caption{Scenario C: A user interacting with an AI Travel Planner.}
    \label{fig:four}
    \Description{A three-panel storyboard follows a man named David interacting with an AI travel recommendation system in an office setting. In the first panel, David engages with the system over multiple conversational turns, sharing information about his preferences and travel budget. The second panel shows the AI recommending a destination and attempting to explain its reasoning using details from the interaction. In the final panel, David appears satisfied and accepting of the explanation, even though the explanation itself may not necessarily be accurate.}
\end{figure*}

\section{Toward Human-Centered Conversational Explainable AI}

The scenarios above reflect how explanations are increasingly encountered in practice, unfolding through dialogue over time, shaped by conversational history and evolving context. In such settings, examining the conversational properties of explanations becomes a natural next step. \textcolor{black}{In this provocation, we introduce HC2XAI as a complementary direction for  existing conversational XAI efforts, and} as a human-centered framing that foregrounds the interactional and conversational dynamics through which explanations are experienced in practice. In this section, we outline the key dimensions relevant for inquiry in this space, each paired with a provocative question intended to open, rather than close inquiry. 

\textbf{The unit of explanation.} In conventional XAI, the explanation is a discrete output: a heatmap, a summary, a set of rules. In conversation, explanation is distributed across turns, packaged with non-explanatory utterances around it. It can unfold slowly, be interrupted, requested again in different words or repaired when it fails to land. The \textit{unit} of an explanation in conversation is not the system's single static output but the interactional sequence that produced and refined it. Mathur et al. \cite{mathur2025sometimes} characterize this as “explanations as conversational exchanges, rather than a conversational dead-end”. The question here then becomes: \textit{How do we study, evaluate and design for explanation as a conversational exchange and process rather than a static conversational outcome?}

\textbf{Persona and voice.} For several years, the CUI community has been engaged with questions of system and conversational persona centered around who the system appears to be (including examination of metaphorical understanding \cite{desai2023metaphors, desai2024examining}) and how its personality and voice shape user experience and trust \cite{rheu2021systematic}. While this work has largely been focused on overall interactions and not specifically on explanations, we see that in a conversational system, persona is not separable from how an explanation is delivered and received. If the same factual content in an explanation is delivered by an agent designed to be warm and cooperative versus one designed to be authoritative and direct, does the user's comprehension, trust calibration or willingness to challenge the AI’s explanation change? \textit{What and under which contexts is the right "presentation" for an AI to explain itself, and what are the implications of that choice for equity and access?} 

\textbf{The user as a conversational partner.} HCXAI has made important strides in characterizing the user: their expertise, goals, mental models, etc. But in a conversational exchange, akin to the explanation, the user is also not a static profile to be queried. In everyday conversational exchanges, the user becomes a dynamic interlocutor whose needs shift across turns and contexts. In this setting, dimensions identified by \citet{Doyle_Clark_Cowan_2021}, such as perceptions of partner competence, dependability, human-likeness, and perceived cognitive flexibility become increasingly relevant. A user who begins a conversation curious and engaged may become frustrated, disengaged or skeptical mid-explanation, and a user who seemed to understand the AI initially may eventually reveal, maybe three turns later, that they want more information. The question then becomes: \textit{How do we build explanation systems that treat users as conversational partners with evolving, in-situ needs, rather than as profiles to be matched to pre-designed explanations?}

\textbf{Evaluation methodology.} The dominant paradigm for evaluating XAI treats explanations as isolated stimuli, asking participants to rate them or use them for decision-making. While well-suited for static, artifact-based explanations, this approach falls short in capturing the dynamics of conversation. An explanation delivered mid-exchange, particularly after a frustrating interaction, carries a different weight than the same explanation encountered in isolation. In such settings, traditional metrics such as reliability and trust remain necessary but insufficient. \textcolor{black}{Recent work has also called for XAI evaluations that move beyond productivity and accuracy-oriented measures toward broader human-centered outcomes, including well-being, trust calibration and sustained engagement \cite{cox2025reflecting, rong2023towards}.} As explanations become embedded in conversational interactions, dimensions from CUI research \cite{Wei_Kim_Kuzminykh_2023}, including likeability, empathy, and perceived responsiveness, may also shape how explanations are received and acted upon, prompting the question: \textit{How do we design evaluation methodologies that would allow us study explanations in situ,  embedded within real conversational contexts and sensitive to the dynamics that precede and follow it?}

\textbf{Appropriateness and accuracy.} While XAI evaluation has focused on accuracy, faithfulness, and comprehensibility, conversational explanations must also be \textit{appropriate}, shaped by situational context, relationship, and the affective dynamics of the exchange. A technically correct explanation delivered too bluntly, too early, or too late may undermine rather than support user understanding and trust, a central goal of explanation. This raises questions of when and how explanations should be surfaced, echoing CUI work on proactivity across routine and high-stakes scenarios, where timing, framing, and contextual sensitivity are critical \cite{Zargham_Reicherts_Bonfert_Voelkel_Schoening_Malaka_Rogers_2022}. In this light, \textit{what would it mean to incorporate conversational appropriateness into the criteria by which AI explanations are designed and evaluated?}

\textcolor{black}{While many of these questions remain open, emerging design directions for HC2XAI may include explanations that adapt conversationally across turns, systems that dynamically adjust tone and pacing based on user responses, effective mechanisms for conversational repair when explanations fail to land, and interaction designs that make uncertainty and confidence legible without disrupting conversational flow. The vision for HC2XAI, as articulated, focuses on the design of explanations as evolving conversational processes that shape how AI systems communicate their reasoning to users with varying needs and requirements for explanations.}

\section{Conclusion}

\begin{comment}
A well-known principle in communication theory, associated with Watzlawick et al. \cite{watzlawick2010human}, is that one cannot simply not communicate, and every communicative act that might not even include words, such as silence, register, timing, also carries meaning in the conversation. The implication of this idea for explanation design is recognizing that the way a Conversational AI system delivers an explanation is always communicating something, whether or not it is explicitly designed to do so. A system that responds to \textit{"why did you say that?"} with an impersonal bulleted list is communicating something about the nature of the interaction, and its transactional character (perhaps even its indifference to the relational context), even if that was not an intended design consideration.
\end{comment}

Research in XAI has largely prioritized getting the explanatory content right, interrogating critical issues in complex AI systems such as high-stakes decision-making, accountability, and opening the “black box” of otherwise opaque models \cite{chahar2025research, zerilli2022explaining}. But as explanations move into everyday conversation, we inherit all the concerns and examinations that conversational dynamics bring into the picture: the expectation of interactional responsiveness, the possibility of repair and the social weight of tone and timing. Increasingly, designing for conversational XAI means designing for all of these nuances, in addition to the informational content of the explanation.

Fortunately, the CUI community has spent years developing the conceptual and empirical tools to take conversation seriously as a design space. In parallel, HCXAI has spent years developing the conceptual and empirical tools to take the “human” seriously as the recipient of an explanation from an AI system. What we need now is to bridge these fields together to produce a cohesive structure for Human-Centered Conversational Explainable AI. The intellectual resources for it are largely in place already, but what is missing is the intention to bring them together. This is not without challenges, and designing conversational explanations that are both technically grounded and interactionally appropriate will require expertise across NLP, CUI, HCI and conversation analysis, and their evaluation will demand more ecologically valid methods than static paradigms allow. But these are not reasons to avoid the problem. They are reasons to start now, while the systems are still new enough to be shaped.

\begin{acks}
    
GPT-5.4 was used to create the initial versions of Figures 1, 2, 3 and 4, which were further edited and refined by the authors.

\end{acks}

\bibliographystyle{ACM-Reference-Format}
\bibliography{cites}

\end{document}